\makeatletter
\declare@file@substitution{revtex4-1.cls}{revtex4-2.cls}
\makeatother

\documentclass{aastex631}
\usepackage{CJK}

\usepackage{newtxtext,newtxmath}
\usepackage[T1]{fontenc}
\usepackage{ae,aecompl}
\usepackage{graphicx}	
\usepackage{amsmath}	
\usepackage{amssymb}	
\usepackage{xcolor}

\newcommand{\br}{\mbox{$G_{BP}-G_{RP}$}}
\newcommand{\teff}{\mbox{$T_{\rm eff}$}}

\received{}
\revised{}
\accepted{}
\submitjournal{ApJ}

\shorttitle{UCDs in the LAMOST}
\shortauthors{Xiang et al.}

\begin{document}
\begin{CJK*}{UTF8}{gbsn}
\title[UCDs in the LAMOST]{Magnetic activity of ultracool dwarfs in the LAMOST DR11}

\correspondingauthor{Yue Xiang, Shenghong Gu}
\email{xy@ynao.ac.cn, shenghonggu@ynao.ac.cn}

\author{Yue Xiang}
\affiliation{Yunnan Observatories, Chinese Academy of Sciences, Kunming 650216, China}
\affiliation{Key Laboratory for the Structure and Evolution of Celestial Objects, Chinese Academy of Sciences, Kunming 650216, China}
\affiliation{International Centre of Supernovae, Yunnan Key Laboratory, Kunming 650216, China}

\author{Shenghong Gu}
\affiliation{Yunnan Observatories, Chinese Academy of Sciences, Kunming 650216, China}
\affiliation{Key Laboratory for the Structure and Evolution of Celestial Objects, Chinese Academy of Sciences, Kunming 650216, China}
\affiliation{School of Astronomy and Space Science, University of Chinese Academy of Sciences, Beijing 101408, China}

\author{Dongtao Cao}
\affiliation{Yunnan Observatories, Chinese Academy of Sciences, Kunming 650216, China}
\affiliation{Key Laboratory for the Structure and Evolution of Celestial Objects, Chinese Academy of Sciences, Kunming 650216, China}
\affiliation{International Centre of Supernovae, Yunnan Key Laboratory, Kunming 650216, China}

\begin{abstract}
Ultracool dwarfs consist of lowest-mass stars and brown dwarfs. Their interior is fully convective, different from that of the partly-convective Sun-like stars. Magnetic field generation process beneath the surface of ultracool dwarfs is still poorly understood and controversial. To increase samples of active ultracool dwarfs significantly, we have identified 962 ultracool dwarfs in the latest LAMOST data release, DR11. We also simulate the Chinese Space Station Survey Telescope (CSST) low-resolution slitless spectra by degrading the LAMOST spectra. A semi-supervised machine learning approach with an autoencoder model is built to identify ultracool dwarfs with the simulated CSST spectra, which demonstrates the capability of the CSST all-sky slitless spectroscopic survey on the detection of ultracool dwarfs. Magnetic activity of the ultracool dwarfs is investigated by using the H$\alpha$ line emission as a proxy. The rotational periods of 82 ultracool dwarfs are derived based on the Kepler/K2 light curves. We also derive the activity-rotation relation of the ultracool dwarfs, which is saturated around a Rossby number of 0.12.
\end{abstract}

\keywords{Stellar activity (1580), Starspots (1572), M dwarf stars (982), Space telescopes (1547)}

\section{Introduction}

At the coolest end of the main sequence, ultracool dwarfs (UCDs) include the lowest-mass stars and substellar objects. Their spectral types are M7 or cooler, and they have mass less than 0.1 M$_{\odot}$ and effective temperature lower than 2700 K \citep{kirkpatrick1997}. They have the highest abundance in the solar vicinity and are important for the studies on stellar, planetary and galactic evolution. UCDs are the bridge between stars and giant planets \citep{burrows2001} and strongly influence the atmospheric evolution of around planets through their intense high-energy radiation \citep{bourrier2017}. Due to their faint brightness, spectroscopic observations of UCDs are almost limited to the solar neighbourhood. UCDs have been identified using data from several surveys, such as the Baryon Oscillation Sky Survey \citep{schmidt2015} and the Javalambre Photometric Local Universe Survey \citep{masbuitrago2022}. The machine learning approach was widely used in identification and characterization of UCDs from spectroscopic and photometric data \citep{marocco2019,avdeeva2024}.

Magnetic field plays a key role in formation and evolution of low-mass stars. The magnetic field on the Sun and Solar-type stars is believed to be generated by the dynamo process taken place within the tachocline layers between the radiation cores and convection envelopes. For a star with mass less than about 0.35 $M_{\odot}$ \citep{chabrier1997}, its interior is fully convective so that there is no tachocline layer, the magnetic field generation mechanism is thought to be very different from the Sun-like stars. However, several studies demonstrated that the activity-rotation relation on low-mass stars is very similar to that of the Sun-like stars, which makes the role of the tachocline layer in the dynamo process controversial \citep{wright2016,newton2017}. The magnetic activity cycle, such as the 11-year solar cycle, is common on Sun-like stars, but is not expected for a fully convective low-mass star at the end of the main-sequence from numerical simulations.

UCDs are attracting increasing attention for exoplanet detection, especially for habitable terrestrial planets \citep{gillon2017,zieba2023,gillon2024}. Given their very low masses, small rocky planets are expected to be more common around them \citep{alibert2017}. Moreover, since the habitable zones locate much closer to the very-low-mass stars than those of hotter stars, potential planets have shorter orbital periods and are more detectable \citep{montgomery2009}. The magnetic field of UCDs plays a very important role on the habitability of planets around them, since they frequently show high level (super)flare phenomena \citep{ram1,ram2,ram3}. \citet{paudel2019} detected frequent superflares on three UCDs with spectral types M7--M9.

The Large Sky Area Multi-Object Fiber Spectroscopic Telescope (LAMOST or Guoshoujing Telescope; \citealt{lamost}) is a 4-m reflecting Schmidt telescope located at the Xinglong station of National Astronomical Observatories of China. With a large field of view (FoV) of 5\degr and 4000 fibres at the focal plane, it is performing one of the most efficient spectroscopic surveys, to collect low-resolution and mid-resolution spectra. \citet{wang2022} had identified 734 UCDs from the LAMOST DR7. They used a spectral type threshold of M6, which is slightly earlier than the traditional definition, to ensure the inclusion of young brown dwarfs, which are much brighter and hotter than their older counterparts. There are 577 M6, 128 M7, 16 M8, 3 M9, and 1 L0 dwarfs in their samples. They also found nine UCDs that were misclassified by the LAMOST as giant stars. 77 out of 734 UCDs show lithium absorption lines in their spectra, which may indicate them in early evolutionary stages.

The Chinese Space Station Survey Telescope (CSST) is a space-bone telescope with a 2-meter aperture and a FoV of 1.1 deg$^{2}$. It will perform a all-sky photometric and spectroscopic survey in the near future, with its multi-band imagers and slitless spectrographs \citep{gong2019}. With the limiting magnitude of about 24 for regular survey and 26 for deep field, it will be among the most efficient surveys for faint stars, like UCDs. In the previous study, we demonstrated the ability of the detection and characterization of stellar magnetic activity with the very-low-resolution slitless spectra by using the H$\alpha$ line emission as a proxy \citep{xiang2022}. Here, we expand our studies to investigate the ability of observations for UCDs with the CSST. From the LAMOST DR11 V1.0, more new samples of UCDs are identified both through the cross-matching with the existed catalogs and a semi-supervised model for the classification of dwarfs and giants from the degraded LAMOST spectra which mimic the slitless spectra of the CSST, and then we investigate the magnetic activity of these UCDs. In Section 2, we shall describe the sample selection procedure for new candidates from the LAMOST DR11 and the semi-supervised classification with an autoencoder model. In Section 3, we analysis the magnetic activity with the H$\alpha$ emission lines for the low-mass stars and the activity-rotation relation. In section 4, we discuss our results and also provide a perspective on the CSST era. Finally, we summarize the present work in Section 5.

\section{Sample selection}

The LAMOST DR11 V1.0 has collected a total number of 11,939,296 low-resolution spectra from 11-year observations, whose resolution is about 1800 at 5500\AA. It provides 898,350 spectra of M stars as well as the stellar atmospheric parameters, including effective temperature (\teff), surface gravity ($\log g$) and metallicity ([M/H]), derived by using the LAMOST Stellar Parameter Pipeline for M stars (LASPM, \citealt{du2021}). The LASPM performs the template matching to a grid of model atmosphere spectra for the low-mass stars from the BT-Settle CIFITS2011 \citep{allard2011,allard2012} to find best-fit stellar atmospheric parameters. In the previous study to select UCDs, \citet{wang2022} found that the \teff\ values derived by the LASPM are systematically 200K higher than those in literature.

Since many targets in the LAMOST survey have multiple observations and there is no need to duplicate the identification of spectra from the same star, we use the LAMOST LRS multiple-epoch catalog of DR11 to construct a database of M stars with spectra types later than M6 from the LAMOST LRS catalog of M stars, where we only keep the spectra with the highest signal-to-noise ratio in the z band (SNRz) for the targets with multi-epoch observations. We have also removed the targets with the SNRz lower than 2 from the samples, which are insufficient for the following analysis.

\subsection{Cross-matching with existed catalogs}

Since there are many existed catalogs of UCDs from all-sky surveys, we begin our identification from a cross-matching. We have searched the late-M stars in the LAMOST survey to select candidates within the target list of the SPECULOOS \citep{sebastian2021}, ultracool dwarf sample in the Gaia DR3 \citep{sarro2023}, and the known low-mass and T Tauri stars in the SIMBAD \citep{wenger2000}, using a circular radius of 5 arcseconds. Then we have visually inspected each spectrum with the stellar spectral typing tool in the Python, PyHammer \citep{kesseli2017,roulston2020}, to ensure that the targets belong to dwarfs with spectral types cooler than M6. As a result, 266 targets are identified in the sample.

We have also used the PyHammer to check the classes of the ultracool dwarf sample of \citet{wang2022} to get more precise spectral types for the following analysis, since the spectral type is a very important factor for studying stellar magnetic activity.

\subsection{Semi-supervised identification}

\subsubsection{The CSST spectrum simulation}

The CSST will take a deep all-sky survey, and collect massive multi-band photometric and slitless spectroscopic data, which will provide a good opportunity to the ultracool dwarf science by discovering new low-mass objects and characterizing them. Given the huge data in the future, it is impossible to visually inspect all spectra to classify the UCDs. On the other hand, the resolution of the CSST slitless spectra is very low as 200-250. Thus, we try to develop a method with the machine learning approach for the identification of dwarfs from late-M stellar spectra with a low resolution of 250. 

\begin{figure}
\centering
\includegraphics[width=0.45\textwidth]{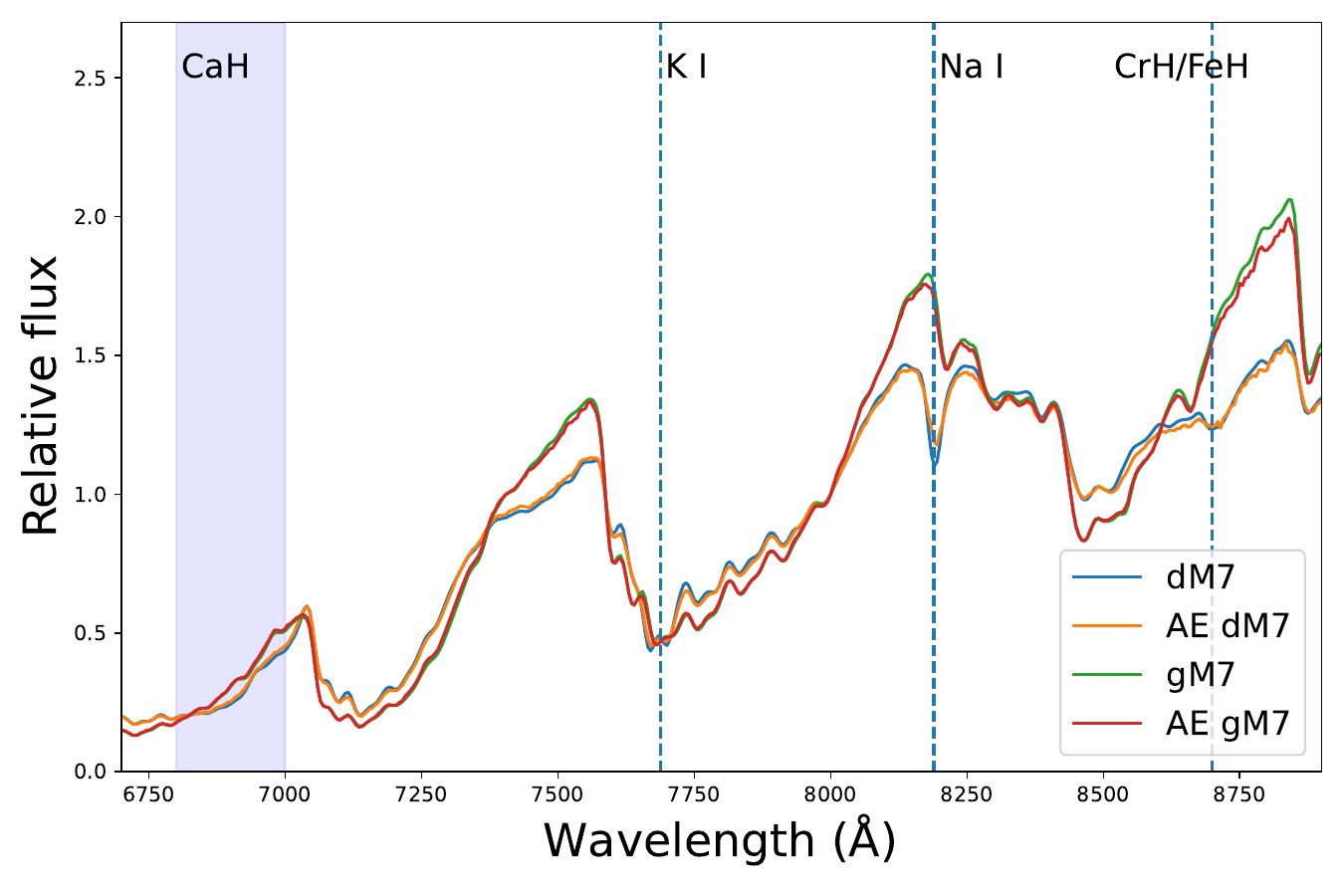}
\includegraphics[width=0.45\textwidth]{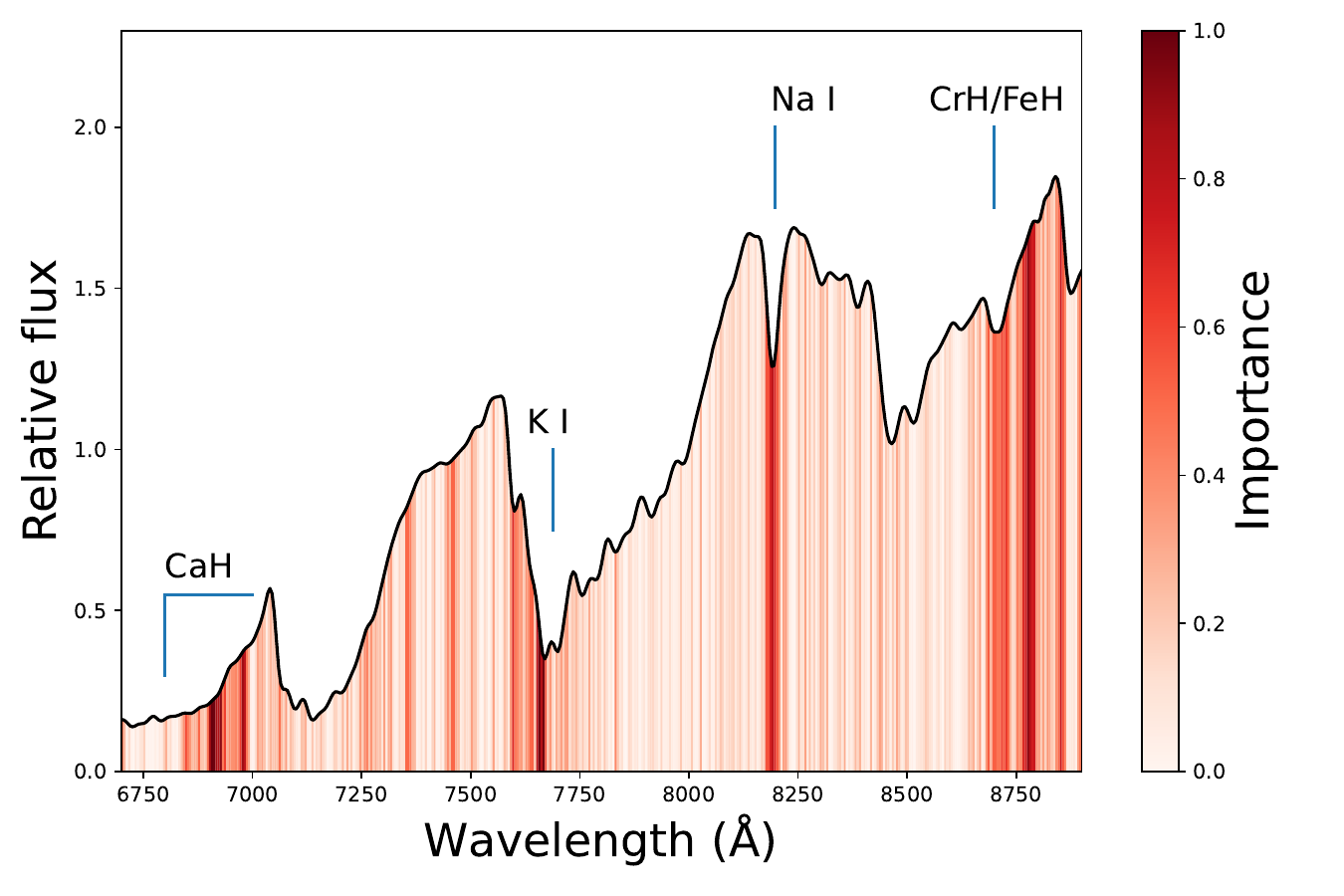}
\caption{The left panel shows examples of the simulated CSST spectra of M7 dwarf and giant stars, as well as the reconstructed results from the autoencoder. The right panel shows the saliency map for the neural networks on a dwarf spectrum.}
\label{fig:vae}
\end{figure}

In order to mimic the CSST slitless spectra, we have convolved a Gaussian profile to the LAMOST spectra to degrade the spectral resolution \citep{xiang2022}. We normalise the flux by dividing the mean value within a 40\AA\ window centering 8000\AA, for easier comparison between spectra of different stars. For the identification task, we only use the wavelength range from 6700 to 8900\AA. The spectra below 6700\AA\ have lower SNRs and very different H$\alpha$ profiles depending on magnetic activity. And the LAMOST spectra in the range between 8900 and 9000\AA\ are often absent. The spectra are aligned to a same grid with a pixel incremental of 5\AA\, since the neural networks require an identical input dimension. We show examples of simulated spectra of M7 dwarf and giant in Figure \ref{fig:vae}. The simulated spectra with lower resolution preserve the predominant spectral features of the molecular bands and strong atomic lines, since they cover wide wavelength ranges, CaH at 6800 \AA\ and CrH/FeH at 8700 \AA\ for instance, which are useful to distinguish the spectra of dwarfs and giants. Atomic lines that are important for the classification, like K I and Na I, are shallower due the lower resolution.

\subsubsection{Model training}

The traditional template-matching method, like that used by the LASPM, is lower efficient and more sensitive to the abnormal spectra, especially for the coolest ones of the main sequence objects due to low SNRs. For instance, \citet{wang2022} only confirmed 3 out of a total of 230 M9 catalog dwarfs and another 3 dwarfs from gM9 subclass. The small number of cool dwarf sample means very few labels for the training, which makes the supervised learning model is insufficient for our task. Hence we have adopted a semi-supervised method to identify more candidates of UCDs with the simulated CSST spectra. Semi-supervised learning is becoming increasingly popular in astrophysical studies (e.g. \citealt{villar2020}), where labeled data are limited but unlabeled data is much more abundant.

\begin{figure}
\centering
\includegraphics[width=0.8\textwidth]{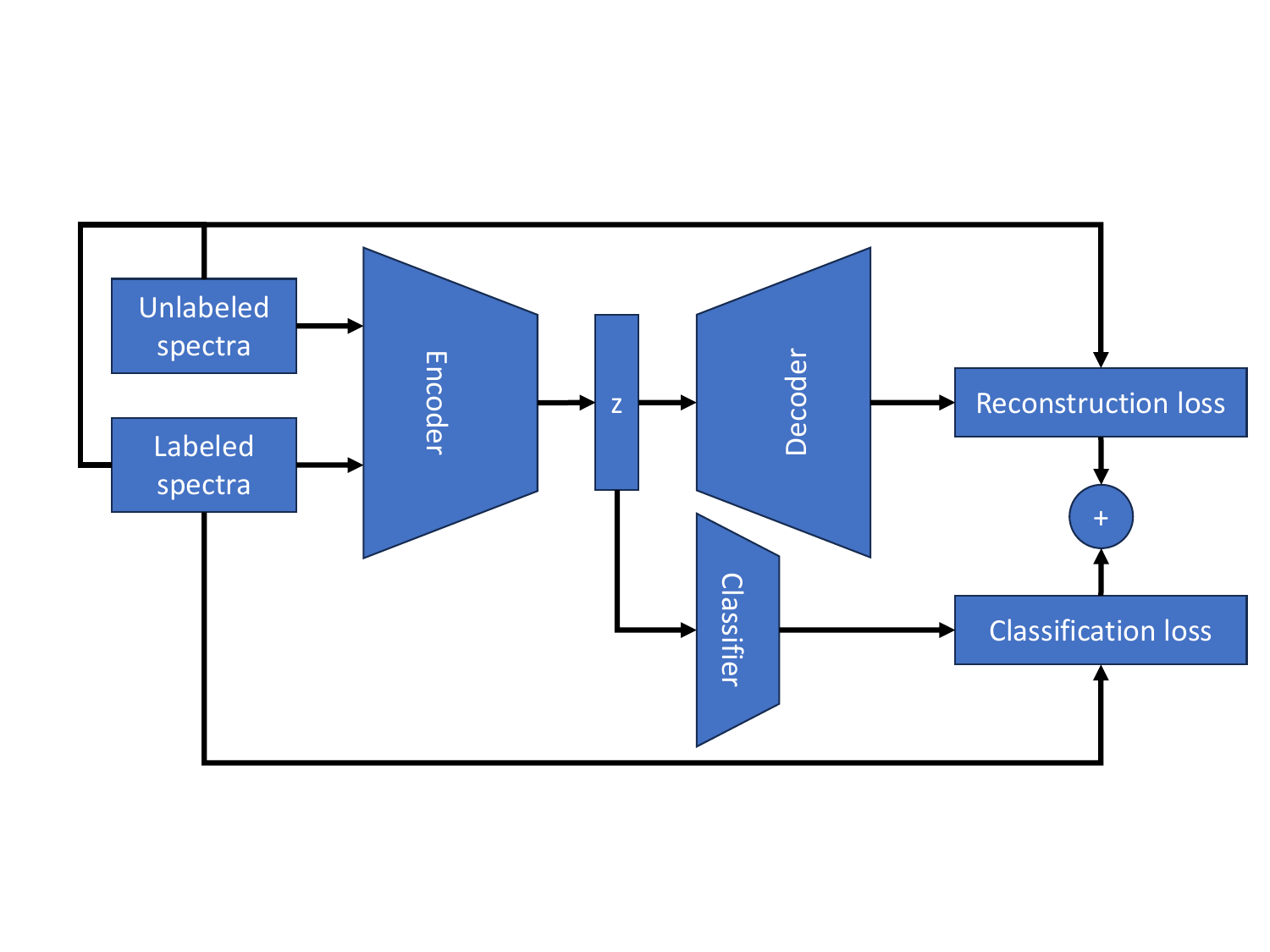}
\caption{Structure of the semi-supervised classification model with an autoencoder.}
\label{fig:ssae}
\end{figure}

Autoencoder is one of the most popular unsupervised models for the representation learning and generative tasks \citep{autoencoder}. It is usually a bottle-neck shaped neural network, which is commonly used for feature extraction, dimension reduction, anomaly detection, and so on. We have built a autoencoder model to learn the representation ability from both of the labeled and unlabeled spectra to extract the features from the spectra, and we connect a neural network of classifier with an activation function of Sigmoid to the latent layer to classify the labeled spectra. The model is constructed with the Python deep learning framework PyTorch \citep{pytorch}. The spectra are standardized with the RobustScalar in Scikit-learn \citep{sklearn} to ensure the deviation on each wavelength point has a similar scale. The reconstruction loss is the mean squared error (MSE) between the input and output spectra, while the classification loss is the cross entropy. Then the total loss is the summation of the reconstruction and classification loss. In figure \ref{fig:ssae}, we show the structure of the neural network model and the working flow of the training process. The encoder consists of three fully connected layers with 512, 256, and 128 nodes, while the decoder has a similar structure but with the layers in a reverse order. The dimension of the latent is 32. The activation function of ReLU \citep{relu} is used for all layers except for the last one.

We have selected the spectra of dwarfs in \citet{wang2022} and the giants classified by the LASPM with high SNRs in the LAMOST DR7 dataset as the labeled dataset and all spectra in the LAMOST DR7 as the unlabeled dataset for the training. The radial velocities of the spectra can be much more diverse in the real data than those in the training set. We manually shift the training spectra with random velocities in range between -300 and 300 km/s to augment the data. We also add random Gaussian noises in the input data, but calculate the reconstruction error with the original data. The denoising autoencoder is more robust to the noisy data.

\subsubsection{Results}

\begin{figure}
\centering
\includegraphics[width=0.5\textwidth]{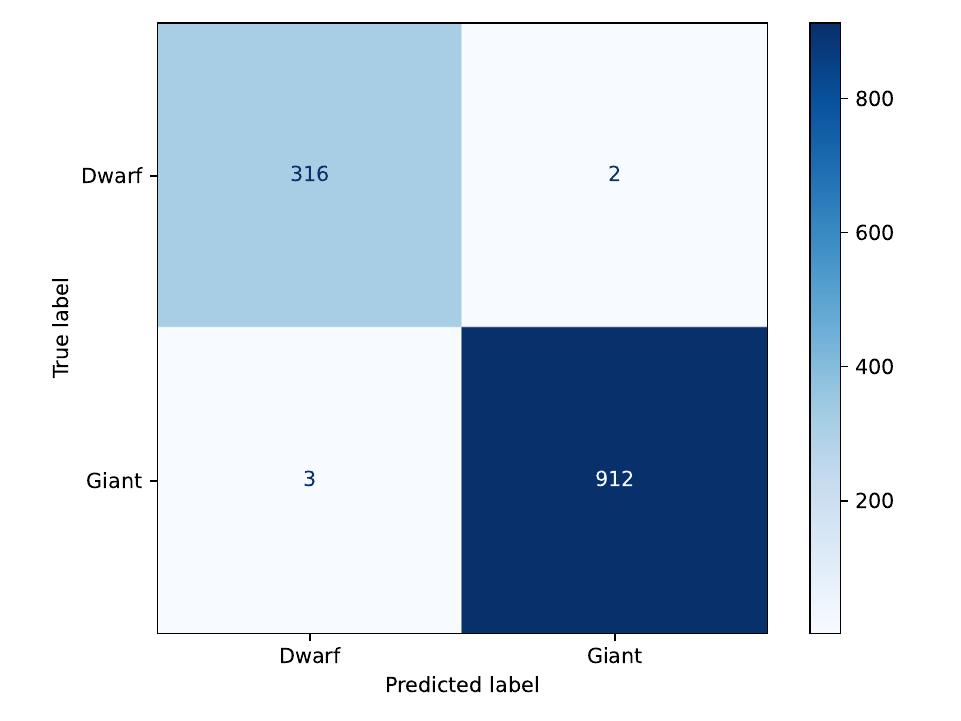}
\caption{Confusion diagram for the classification on the test data set.}
\label{fig:cm}
\end{figure}

We show the simulated CSST slitless spectra along with the reconstructed ones by the autoencoder in Figure \ref{fig:vae} as well. We construct a test dataset containing the cross-matched UCDs's spectra and the duplicated observations of targets of \citet{wang2022} in the LAMOST DR11 and giants' spectra newly observed in the LAMOST DR11 to validate the performance of the classification model. We show a confusion matrix in Figure \ref{fig:cm}. The spectra of young candidates are similar to that of the evolved giant stars and may confuse the classification. With the trained model, we have identified 696 UCDs in the LAMOST DR11. Again, we use the PyHammer to visually check the spectral types of the candidates.

Anomaly detection is also important for large surveys to avoid effects on following analyses. From the reconstruction, we can also detect the abnormal spectra from a large dataset. Since the autoencoder model is trained with the normal spectra, it can not perform correct reconstruction on abnormal data and thus results in higher MSEs.

For the stellar spectral classification of dwarfs and giants with visual inspection, we rely on several specific spectral features, such as molecular bands of TiO, CrH and FeH, and atomic lines of Na I and K I doublet \citep{gray2009}. The neural networks extract spectral features automatically rather than manual selection. We derive the saliency map to show which fluxes along the wavelength are important for the neural networks to identify the spectra of dwarfs. The saliency map is the gradient of the classification label to the input \citep{saliency} and it shows how the input fluxes at different wavelength influence the classification result. The resulting saliency map shows that the model has captured the same important features of the atomic and molecular lines as those in the manual classification.

\subsection{Characterization of the sample}

To characterize the colors and kinematic properties of the ultracool dwarf sample, we have cross-matched the positions of the LAMOST targets with the Gaia DR3 \citep{gaiadr3} using an angular radius of 5 arcseconds, and acquired the mean photometric magnitudes of the Gaia $G$, $G_{BP}$, and $G_{RP}$ bands, parallaxes and proper motions. For the duplicated results with the same LAMOST target, we have adopted the one with the largest $G_{BP}-G_{RP}$ color. The absolute $G$ magnitude ($M_{G}$) is calculated following Equation (\ref{eq:mg}) \citep{sarro2023},
\begin{equation}
M_{G} = G + 5  \log_{10}(\varpi)-10
\label{eq:mg}
\end{equation}
where $G$ is the $G$ mean magnitude and $\varpi$ is the parallax in unit of milliarcseconds.

\begin{figure}
\centering
\includegraphics[width=0.45\textwidth]{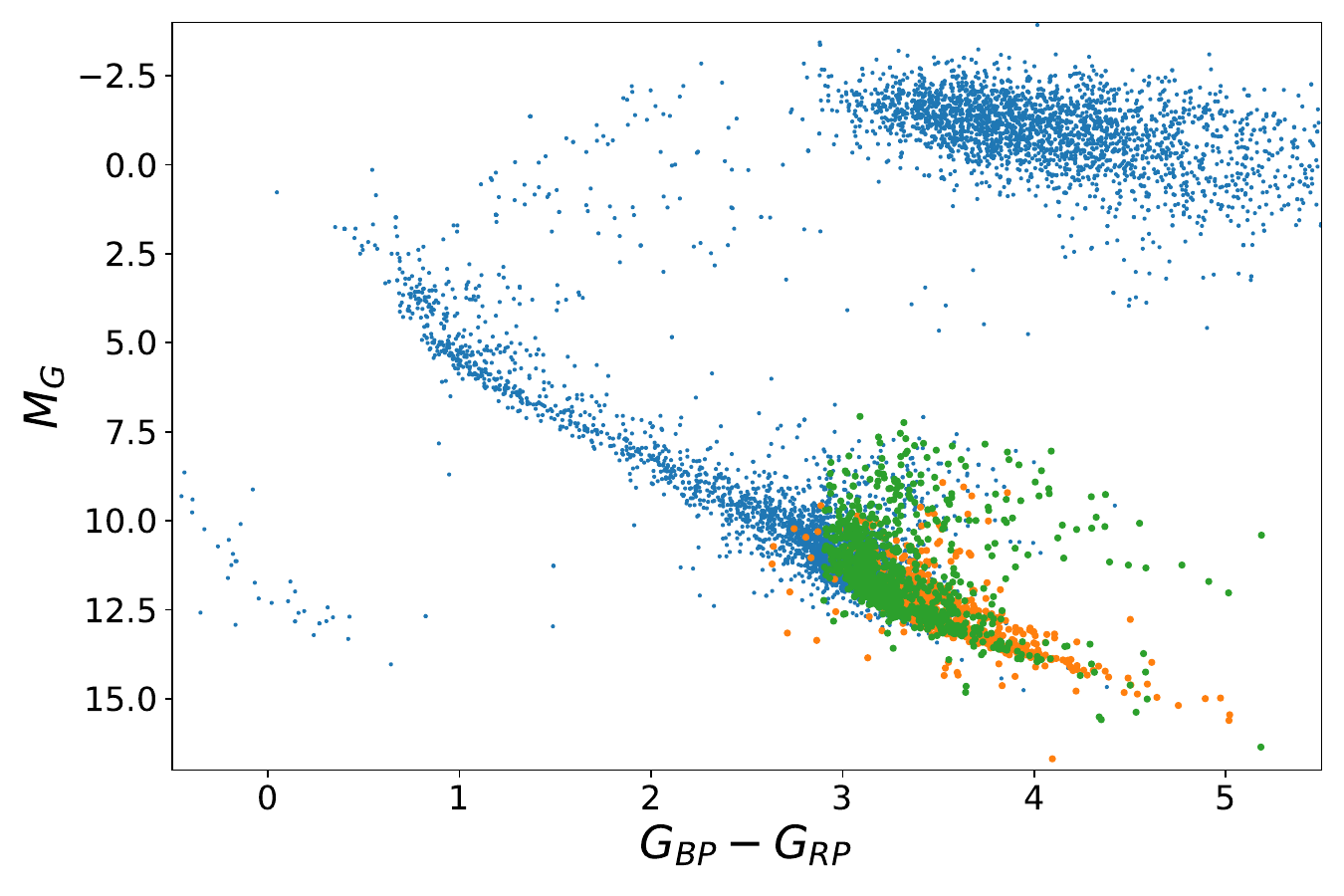}
\includegraphics[width=0.45\textwidth]{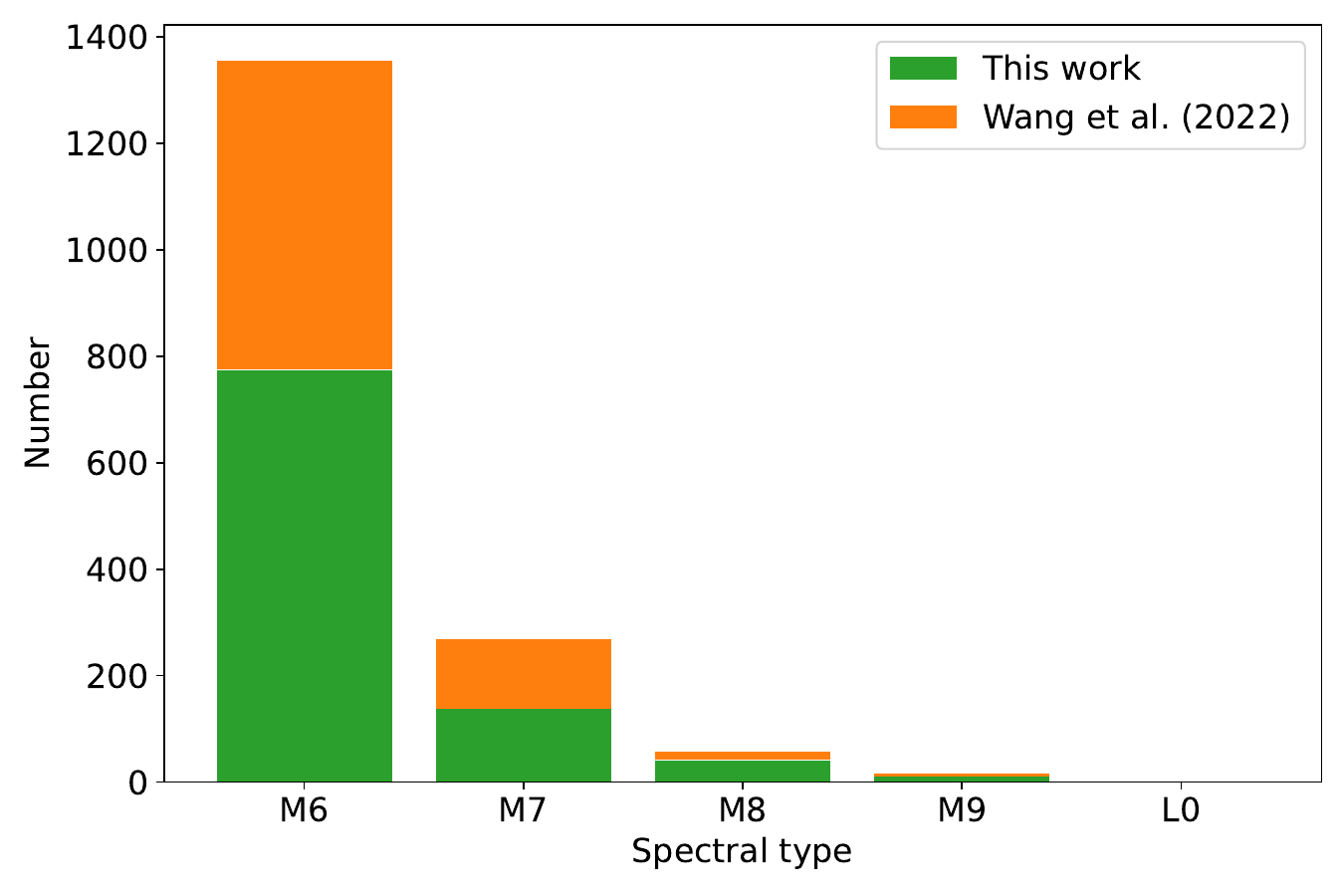}
\caption{Left panel is color-magnitude diagram for the whole M stars (blue), the LAMOST UCDs identified by \citet{wang2022} (orange) and this work (green). Here, we do not take into account the interstellar extinction and reddening. Right panel shows distributions of spectral types for the UCDs identified by \citet{wang2022} (orange) and this work (green).}
\label{fig:hr}
\end{figure}

In Figure \ref{fig:hr}, we show the color-magnitude diagram ($M_{G}$ vs $G_{BP}-G_{RP}$) for the samples identified by \citet{wang2022} and those from this work, along with the entire M stars in the LAMOST DR11 database. In this plot, we do not take into account the interstellar extinction and reddening. Most of the targets are located at distances less than 150 pc, where the extinction is expected to be small. As seen from this figure, the most obvious regions include the low-mass dwarfs and the evolved giants with higher brightness. The UCDs at the early evolutionary stage are located at upper right above the main-sequence and below the giants. The spectral type range of our sample is slightly earlier than the traditional definition of UCDs, which is cooler than M7, to include more lat-M dwarfs. However, all of our sample are well below the fully convective boundary, which is about M4 or 0.35 $M_{\odot}$. The properties of a total of 962 newly identified UCDs from the LAMOST DR11 are listed in Table \ref{tab:ucds}, whose full version is available only in the machine-readable format.

\begin{deluxetable}{cccccccc}
\tabletypesize{\scriptsize}
\tablecolumns{8}
\tablewidth{0pt}
\tablecaption{The properties of the newly identified UCDs.}
 \label{tab:ucds}
\tablehead{
 \colhead{RA}&
 \colhead{Dec}&
 \colhead{LAMOST obsid}&
 \colhead{Gaia source id}&
 \colhead{$M_{G}$}&
 \colhead{$G_{BP}-G_{RP}$}&
 \colhead{Radial velocity}&
 \colhead{Spectral type}\\
 \colhead{(degree)}&
 \colhead{(degree)}&
 \colhead{}&
 \colhead{}&
 \colhead{(mag)}&
 \colhead{(mag)}&
 \colhead{(km/s)}&
 \colhead{}
}

\startdata
0.39965 & 23.29836 & 759602231 & 2848189248602068864 & 13.45 & 3.87 & -46 & M7\\
0.42507 & 57.51464 & 844613079 & 422591663442814080 & 8.48 & 3.20 & -9 & M6\\
0.54728 & 23.76864 & 865314026 & 2848236184005536000 & 11.75 & 3.25 & 7 & M6\\
0.68996 & 47.51442 & 938005057 & 387277682117916800 & 8.20 & 3.30 & -10 & M6\\
0.80209 & 56.81903 & 844606169 & 421054958504899328 & 12.42 & 3.42 & -58 & M6\\
\enddata
\tablecomments{Table 1 is published in its entirety in the machine-readable format.}
\end{deluxetable}

\begin{figure}
\centering
\includegraphics[width=0.5\textwidth]{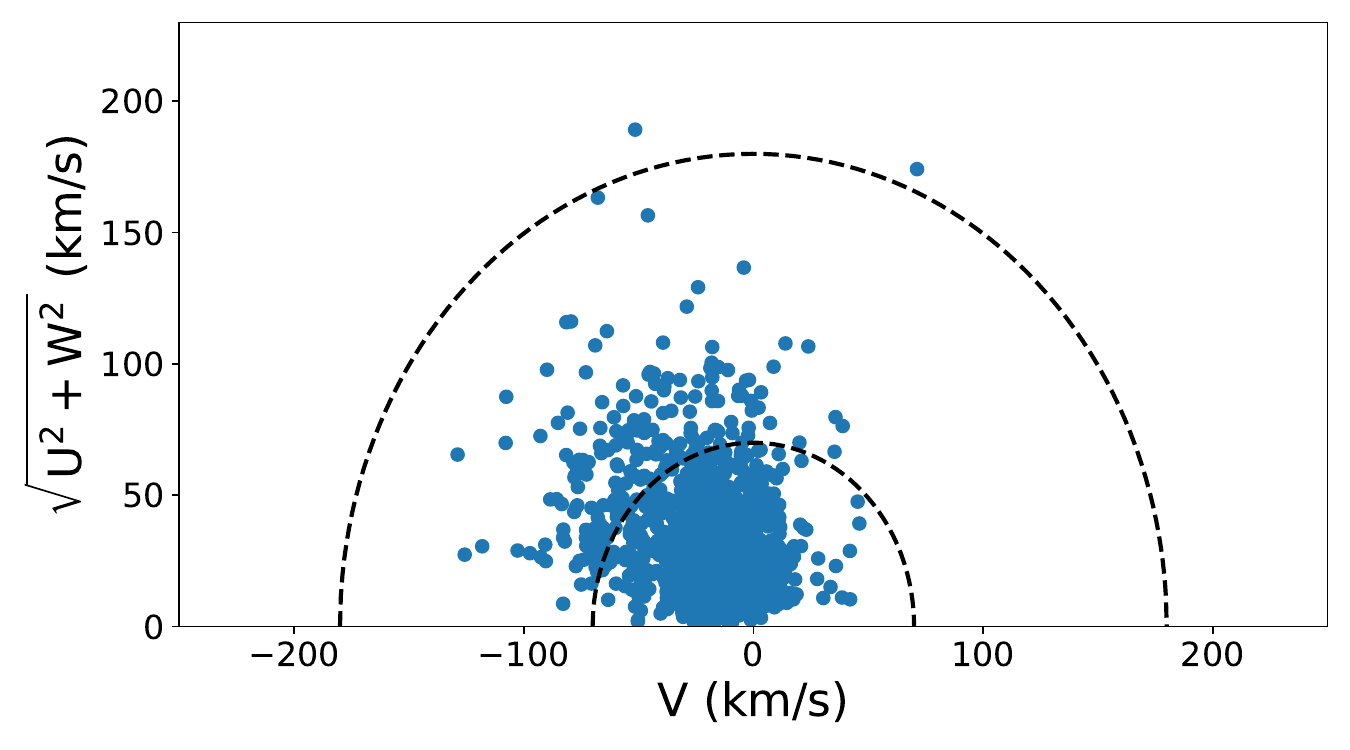}
\caption{Toomre diagram of $UVW$ velocities. The dashed lines show the total velocities of 70 and 180 km/s.}
\label{fig:td}
\end{figure}

We have estimated the radial velocities of the stars from the cross-correlations between the LAMOST spectra of our sample with the PyHammer templates in same spectral types, using pyasl.crosscorrRV in the Python package PyAstronomy \citep{pya}. For the targets with parallaxes larger than five times the errors, we compute the Galactic space velocities ($UVW$) respective to the local standard of rest (LSR), in which $U$ is positive toward the Galactic center, $V$ is positive along the direction of the Galactic rotation, and $W$ is positive toward the north Galactic pole, using the subroutine pyasl.gal\_uvw. The LSR derived by \citet{lsr2010}, ($U_\odot$, $V_\odot$, $W_\odot$) = (11.1, 12.24, 7.25) km/s, is adopted in our calculations. The Toomre diagram of $UVW$ velocities is shown in Figure \ref{fig:td}. The distribution of the stellar motions shows that most of our sample are located in the thin disk and some are located at the disk-halo transition, which may be older stars due to the kinematic evolution. 

\section{Magnetic activity}

\subsection{H$\alpha$ emission}

The H$\alpha$ line is an important indicator for stellar chromospheric activity. We take the equivalent widths (EWs) of the H$\alpha$ line and the measurement errors from the LAMOST M star catalog. The EWs were calculated from a region of 14\AA\ centering the H$\alpha$ line with the continuum estimated from two regions of 6555-6560 \AA and 6570-6575\AA\ \citep{yi2014}, following the method of \citet{west2011}. Note that positive EWs indicate emission lines, while negative values correspond to absorption lines. Therefore, larger positive EWs of the H$\alpha$ line suggest higher activity levels of the stars.

\begin{figure}
\centering
\includegraphics[width=0.45\textwidth]{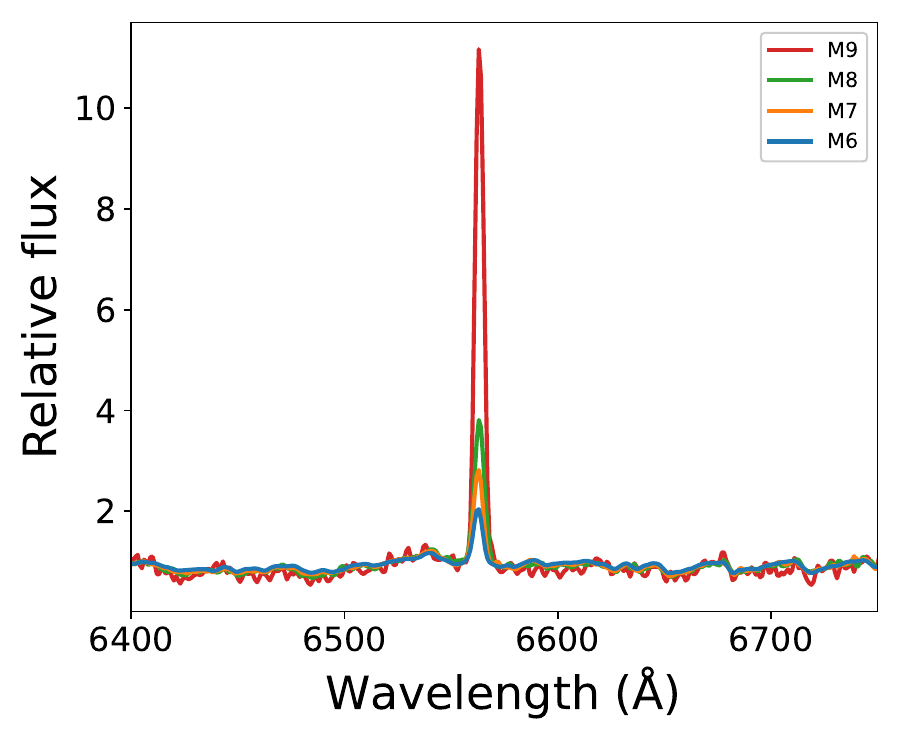}
\includegraphics[width=0.45\textwidth]{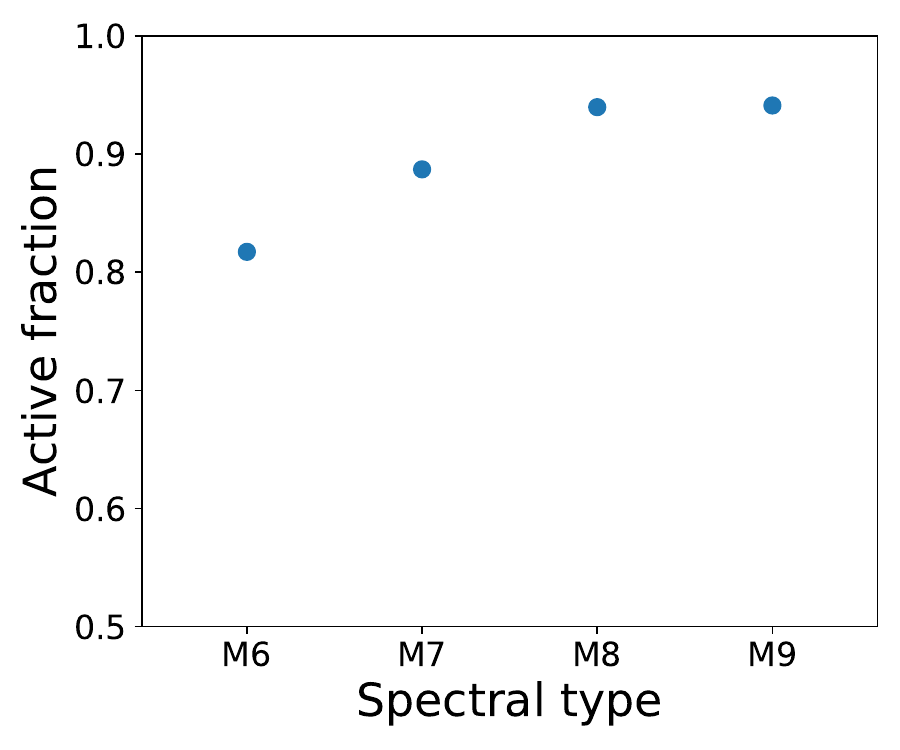}
\caption{Left panel: Normalized mean low-resolution LAMOST spectra around the H$\alpha$ line of newly identified late-M dwarfs of dM6 (blue), dM7 (orange), dM8(green), and dM9 (red). Right panel: The fraction of active UCDs relative to the total sample for different spectral types.}
\label{fig:act}
\end{figure}

Figure \ref{fig:act} shows the examples of active UCDs and their fraction relative to the total sample for spectral types M6-–M9, where the activity is defined as having a EW of the H$\alpha$ line larger than 0.75\AA. Over 80 percent of stars in our sample are active, and the fraction of active stars relative to the total number of each spectral type (active fraction) increases with the decreasing effective temperature from spectral type of M6 to M9, which is consistent to \citet{west2015}. The only one L dwarf in the LAMOST sample is inactive and does not show apparent H$\alpha$ emission, but no conclusive result can be drawn due to the extremely limited sample size.

Since the EWs of the H$\alpha$ line are highly dependent on the stellar effective temperature, it is not a good proxy for the comparison of magnetic activity among stars with different spectral types. $L_{\rm H\alpha}/L_{bol} = \chi_{\rm H\alpha} EW_{\rm H\alpha}$ is more commonly used to represent stellar magnetic activity. Here we have adopted the relation between $\chi_{\rm H\alpha}$ and spectral type from \citet{west2008} to calculate the $\chi_{\rm H\alpha}$ for our sample.

\subsection{Activity-rotation relation}

The stellar rotation is one of the most important factors for the generation of magnetic field. A faster rotation rate increases the efficiency of dynamo, which results in a stronger magnetic field and high level of activity \citep{wright2011,vidotto2014}. The relations of stellar activity to the different rotation profiles can reflect the features of the dynamo processes beneath the stellar surface. The Rossby number, which is the ratio between the stellar rotation period and the convective turnover timescale, is demonstrated to be a better representation for the activity-rotation relation \citep{noyes1984}. The stellar activity-rotation relation has been extensively studied, with all studies revealing a saturation profile: the stellar activity level increases with its rotation, but once the rotation period exceeds a critical point, the activity level remains constant across different bands from X-ray to radio \citep{mclean2012}, which may suggest a saturation of dynamo \citep{reiner2012}. The critical value of the Rossby number is typically around 0.1 \citep{wright2011,reiner2012}.

\begin{figure}
\centering
\includegraphics[width=0.5\textwidth]{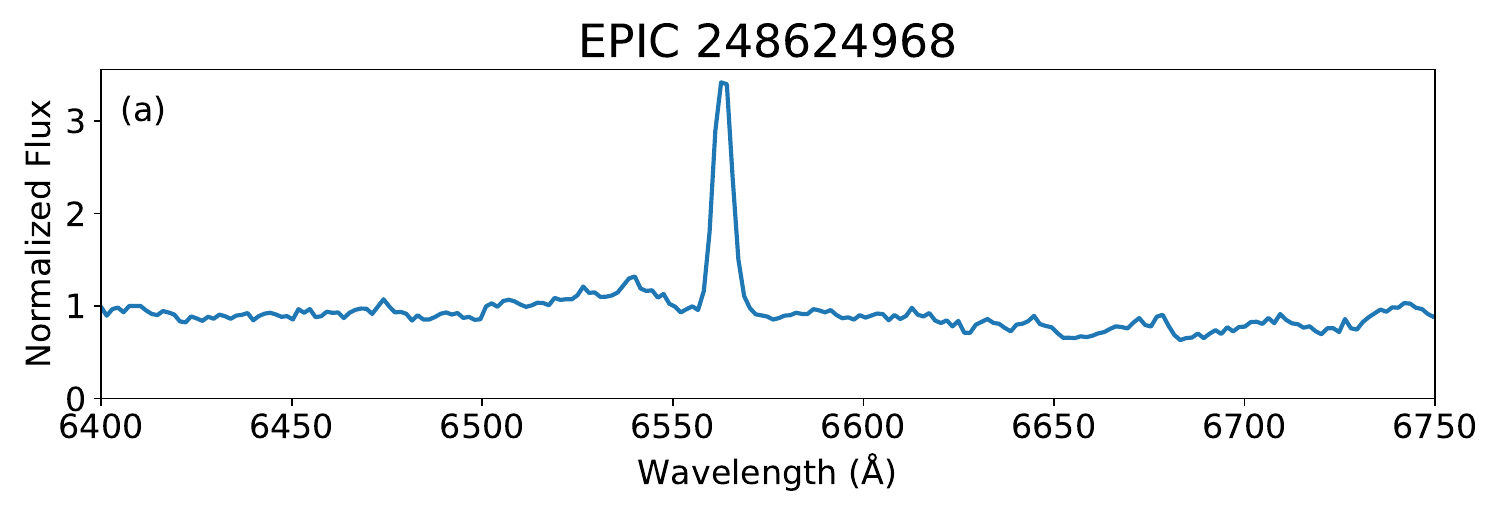}
\includegraphics[width=0.5\textwidth]{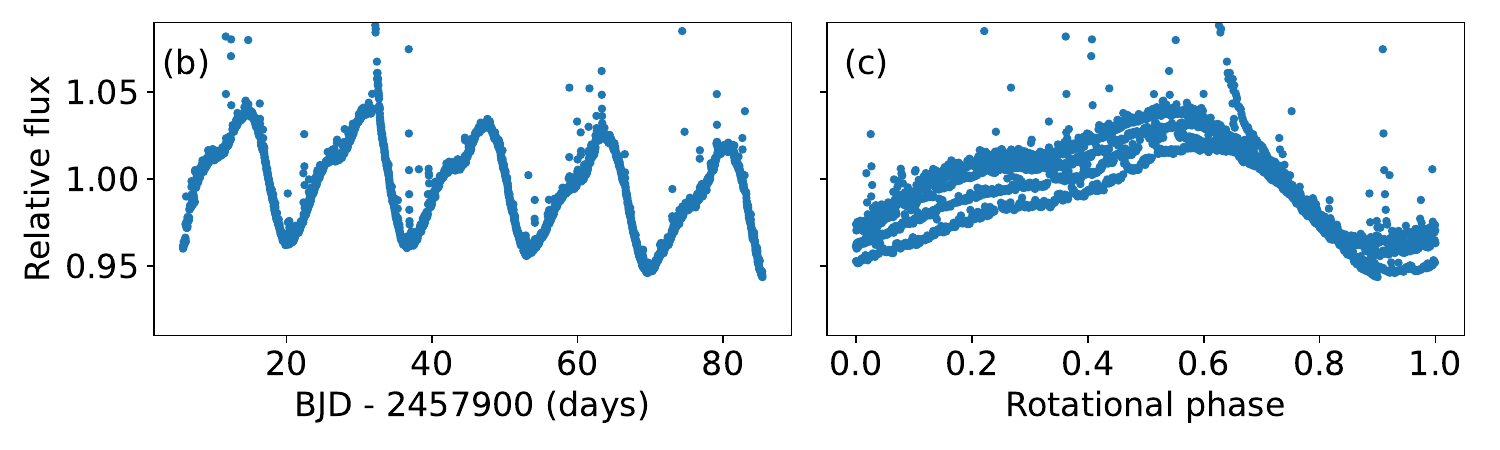}
\includegraphics[width=0.5\textwidth]{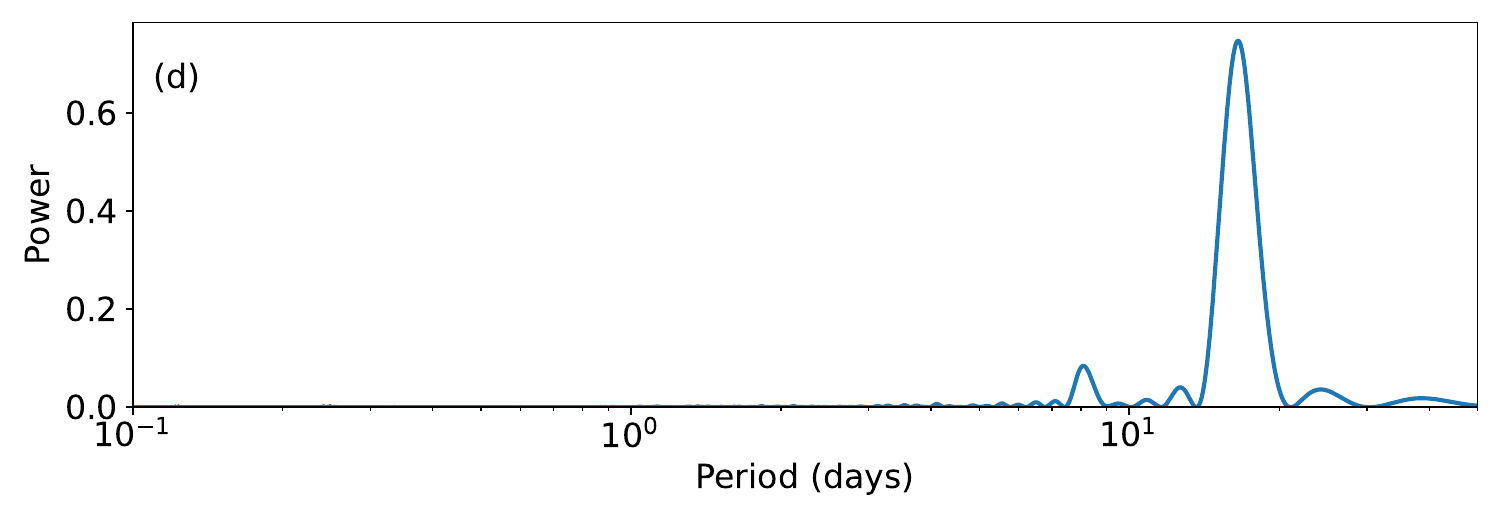}
\caption{Panel (a) shows the LAMOST spectrum around the H$\alpha$ line for EPIC 248624968. Panels (b) and (c) show original and phase-folded K2 light curves, respectively. Panel (d) shows the generalized Lomb-Scargle diagram derived from the K2 photometry.}
\label{fig:freq}
\end{figure}

\begin{figure}
\centering
\includegraphics[width=0.45\textwidth]{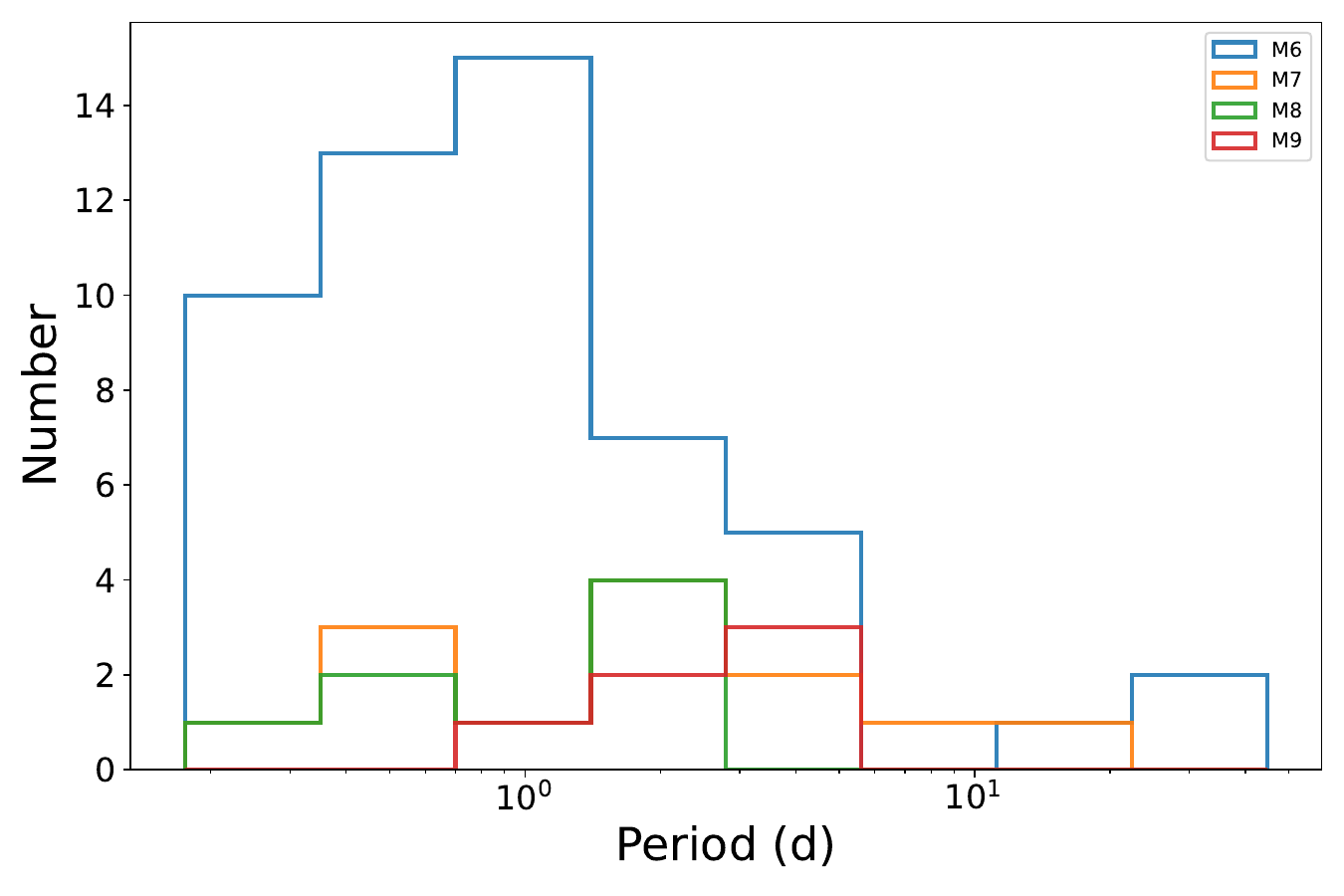}
\includegraphics[width=0.45\textwidth]{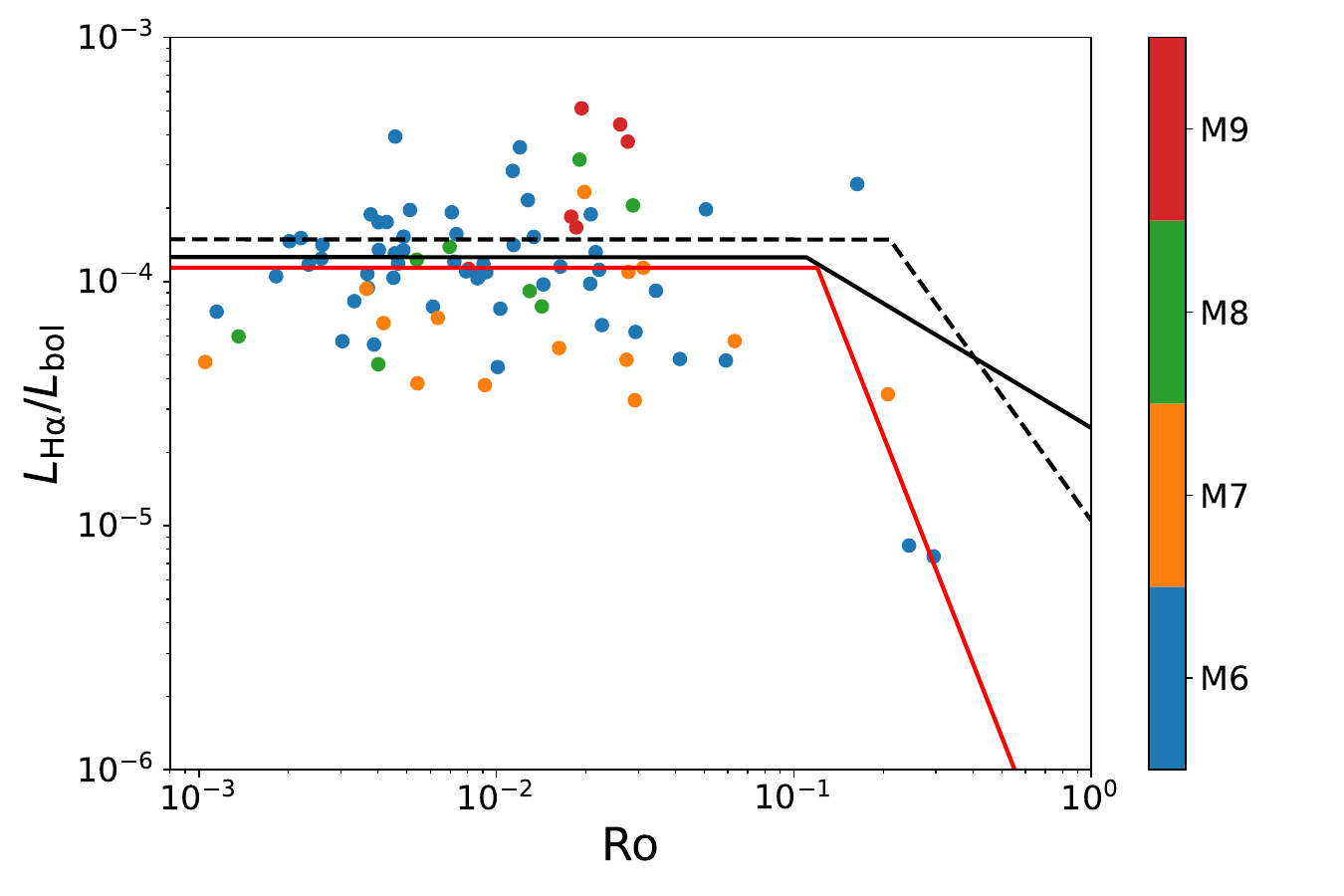}
\caption{Left panel shows the rotational period distribution of targets with different spectral types. Right panel shows $L_{\rm H\alpha}/L_{bol}$ versus the Rossby number, where the red line represents the fitted activity-rotation relation, and the solid and dashed black lines show the relations derived by \citet{douglas2014} and \citet{newton2017}, respectively.}
\label{fig:havsro}
\end{figure}

The stellar rotation can be studied by using high-resolution spectroscopic observations or photometry. Here, we obtain the rotation periods of the UCDs from the Kepler/K2 light curves.  Firstly, we have cross-matched the position of our sample to the Kepler/K2 photometry input catalog within a radius of 5 arcseconds, and acquired the light curves from the server \citep{https://doi.org/10.17909/t98304,https://doi.org/10.17909/t9ws3r}. Only one target has the Kepler photometric data and others have the K2 data. Then we have applied the generalised Lomb-Scargle method \citep{gls} to the Pre-search Data Conditioning Simple Aperture Photometry (PDCSAP) flux, which is corrected for instrumental systematic errors \citep{pdc}, to derive the rotational period. We also visually inspect the phase-folded light curves to ensure that clear rotation-modulated signatures are present. We have hunted 82 ultracool dwarf samples with clear rotational periods. As an example, the original and phase-folded light curve, as well as the Lomb-Scargle periodogram for EPIC 248624968 is shown in Figure \ref{fig:freq}. This star shows a clear rotation period of $16.54 \pm 1.45$ days, which is consistent to the result derived by \citet{reinhold2020}. The light curve showed gradual evolution in shape, which is consistent with the changes caused by evolving starspots. The light curve also reveals frequent flare activities on the star. Furthermore, the star is proven to be very magnetically active by the H$\alpha$ line emission as well, shown in the top panel of Figure \ref{fig:freq}. We show the distribution of the rotational periods of our sample in Figure \ref{fig:havsro}. Most active targets have periods shorter than 5 days, which is consistent with previous studies (e.g., \citealt{newton2017}). We only have four targets with periods longer than 10 days, likely due to the limited time span of the K2 photometric observations.

We adopt the mass values from the Kepler/K2 input catalog \citep{epic}. Some stars has no mass values, and we use the mass values of the stars with the closest \br\ values for the calculations of the Rossby numbers. We have calculated the Rossby number using the relation between it and stellar mass, derived by \citet{wright2018}. All results, including the period, $L_{\rm H\alpha}/L_{bol}$, and Rossby number (Ro), are listed in Table \ref{tab:period}. \ref{fig:havsro}. We fit the activity-rotation relation as the following formula to the data.
\begin{equation}
L_{\rm H\alpha}/L_{bol} = 
\begin{cases}
(L_{\rm H\alpha}/L_{bol})_{sat}, & {\rm Ro} \le {\rm Ro}_{\rm sat}, \\
C{\rm Ro}^{\beta}, & {\rm Ro} > {\rm Ro}_{\rm sat}.
\end{cases}
\end{equation}
and get the results with ${\rm Ro}_{sat} = 0.12^{+0.02}_{-0.04}$ and $\beta = -3.1^{+1.6}_{-0.8}$ shown in Figure \ref{fig:havsro}. The derived ${\rm Ro}_{sat}$ is close to that found by \citet{douglas2014}, but smaller than the value reported by \citet{newton2017}. The fitted $\beta$ is smaller than those derived in these studies, perhaps because our sample is more concentrated in cooler dwarfs. \citet{magaudda2020} found that the slope is steeper in the unsaturated regime for lower-mass stars. However, the small number of targets in the unsaturated regime prevents us from deriving a precise activity-rotation relation. Several stars show much higher levels of the H$\alpha$ emission, which are mainly due to their young ages, because the prominent accretion process also contributes significantly to the H$\alpha$ line emission.

\begin{deluxetable*}{ccccccc|ccccccc}
\tabletypesize{\scriptsize}
\tablecolumns{8}
\tablewidth{0pt}
\tablecaption{The derived activity and rotation information for 82 UCDs in the LAMOST DR11. The last record only has the Kepler photometry, so its Kepler ID is shown in the EPIC column.}
 \label{tab:period}
\tablehead{
 \colhead{RA}&
 \colhead{Dec}&
 \colhead{Obsid}&
 \colhead{EPIC}&
 \colhead{Period}&
 \colhead{$L_{\rm H\alpha}/L_{bol}$}&
 \colhead{Ro}&
 \colhead{RA}&
 \colhead{Dec}&
 \colhead{Obsid}&
 \colhead{EPIC}&
 \colhead{Period}&
 \colhead{$L_{\rm H\alpha}/L_{bol}$}&
 \colhead{Ro}\\
 \colhead{(degree)}&
 \colhead{(degree)}&
 \colhead{}&
 \colhead{}&
 \colhead{(day)}&
 \colhead{}&
 \colhead{}&
 \colhead{(degree)}&
 \colhead{(degree)}&
 \colhead{}&
 \colhead{}&
 \colhead{(day)}&
 \colhead{}&
 \colhead{}
}

\startdata
17.96728 & 2.96469 & 964014060 & 220315502 & 0.52 & 7.89e-05 & 6.12e-03 & 68.46891 & 22.94081 & 680909088 & 247596872 & 1.41 & 9.14e-05 & 1.29e-02\\ 
53.75870 & 23.70989 & 470614137 & 211046195 & 0.22 & 5.97e-05 & 1.36e-03 & 68.47021 & 18.05460 & 1048616102 & 247035365 & 2.59 & 3.27e-05 & 2.92e-02\\ 
53.94274 & 16.56224 & 633508176 & 210572738 & 0.45 & 6.76e-05 & 4.18e-03 & 68.74680 & 17.56057 & 883107061 & 246983144 & 1.60 & 5.35e-05 & 1.62e-02\\ 
54.50813 & 16.84438 & 633508120 & 210591703 & 0.29 & 1.42e-04 & 2.60e-03 & 68.92433 & 22.56988 & 711002027 & 247548866 & 0.69 & 1.39e-04 & 6.96e-03\\ 
56.33321 & 24.47337 & 1086610249 & 211094384 & 0.76 & 1.57e-04 & 7.34e-03 & 68.92517 & 22.87290 & 184801245 & 247588169 & 7.15 & 5.71e-05 & 6.33e-02\\ 
56.60638 & 18.58035 & 712008121 & 210715010 & 0.33 & 1.46e-04 & 2.02e-03 & 68.96708 & 22.91774 & 628801242 & 247593952 & 0.85 & 1.10e-04 & 7.92e-03\\ 
56.62383 & 21.03808 & 783614026 & 210877263 & 0.32 & 1.08e-04 & 3.68e-03 & 69.41139 & 18.70236 & 960906147 & 247103315 & 0.76 & 1.09e-04 & 9.25e-03\\ 
56.73051 & 24.18787 & 66404227 & 211075914 & 0.24 & 1.17e-04 & 2.34e-03 & 69.50350 & 25.98256 & 94315190 & 248023915 & 0.66 & 4.59e-05 & 4.01e-03\\ 
56.93612 & 23.70077 & 1086602232 & 211045643 & 0.45 & 1.19e-04 & 4.67e-03 & 69.58891 & 26.15374 & 628815008 & 248046139 & 2.67 & 2.05e-04 & 2.88e-02\\ 
56.99519 & 23.48452 & 781403010 & 211032304 & 0.32 & 5.52e-05 & 3.88e-03 & 69.64702 & 26.17689 & 721616233 & 248049115 & 4.52 & 1.98e-04 & 5.07e-02\\ 
57.11077 & 23.19165 & 615405067 & 211013604 & 0.46 & 1.35e-04 & 4.03e-03 & 69.74468 & 23.39980 & 628801169 & 247657780 & 0.66 & 1.23e-04 & 5.40e-03\\ 
57.29742 & 18.05444 & 416707089 & 210678008 & 0.18 & 1.51e-04 & 2.20e-03 & 69.93702 & 26.03126 & 628815113 & 248030407 & 3.45 & 1.10e-04 & 2.78e-02\\ 
57.53513 & 25.54873 & 1086604232 & 211156183 & 1.75 & 1.12e-04 & 2.22e-02 & 69.94785 & 26.02801 & 759511023 & 248029954 & 2.96 & 5.13e-04 & 1.93e-02\\ 
57.74792 & 14.23380 & 1070314231 & 210435290 & 0.51 & 1.53e-04 & 4.88e-03 & 70.15416 & 16.63581 & 1048609147 & 246887841 & 1.28 & 3.55e-04 & 1.20e-02\\ 
58.72048 & 26.03017 & 837011035 & 211180726 & 22.13 & 3.46e-05 & 2.07e-01 & 70.26769 & 25.96553 & 184709164 & 248021602 & 2.10 & 1.89e-04 & 2.08e-02\\ 
59.12675 & 24.28852 & 841709143 & 211082433 & 0.34 & 1.88e-04 & 3.78e-03 & 70.45107 & 25.57508 & 759511098 & 247968420 & 2.92 & 3.74e-04 & 2.77e-02\\ 
59.13385 & 22.93013 & 1086607131 & 210997213 & 0.89 & 1.05e-04 & 8.73e-03 & 70.58757 & 25.34288 & 94409023 & 247935696 & 1.89 & 2.33e-04 & 1.98e-02\\ 
59.55993 & 12.62802 & 1070305123 & 210365286 & 0.87 & 1.18e-04 & 9.05e-03 & 71.25125 & 17.27209 & 282912169 & 246953721 & 2.12 & 9.80e-05 & 2.07e-02\\ 
59.95764 & 12.68344 & 1070305079 & 210367469 & 0.46 & 1.97e-04 & 5.12e-03 & 71.95181 & 14.62337 & 951315003 & 246704060 & 2.36 & 1.32e-04 & 2.16e-02\\ 
60.17841 & 14.06798 & 1056004213 & 210427204 & 0.38 & 1.31e-04 & 4.56e-03 & 73.72269 & 20.86945 & 730501051 & 247336308 & 1.64 & 1.15e-04 & 1.64e-02\\ 
61.50168 & 21.87692 & 881110232 & 210930754 & 0.97 & 1.41e-04 & 1.14e-02 & 74.65902 & 16.22240 & 1069914079 & 246846196 & 3.71 & 4.82e-05 & 4.14e-02\\ 
61.89590 & 22.62758 & 170708215 & 210977750 & 0.13 & 4.69e-05 & 1.05e-03 & 75.31666 & 16.14470 & 1069915051 & 246838465 & 1.45 & 9.73e-05 & 1.44e-02\\ 
63.70957 & 24.37690 & 881207171 & 211088189 & 0.45 & 9.42e-05 & 3.70e-03 & 76.69428 & 21.07490 & 939510073 & 247360583 & 0.44 & 3.93e-04 & 4.57e-03\\ 
65.18605 & 18.09983 & 420215156 & 210681303 & 0.96 & 2.84e-04 & 1.14e-02 & 77.90577 & 15.92949 & 843203241 & 246817835 & 5.52 & 4.75e-05 & 5.91e-02\\ 
65.21974 & 17.77821 & 966002088 & 210658027 & 0.36 & 1.75e-04 & 4.28e-03 & 90.56903 & 23.27603 & 760414048 & 202137543 & 0.61 & 7.10e-05 & 6.36e-03\\ 
65.52068 & 19.58011 & 960914071 & 210780956 & 2.89 & 1.14e-04 & 3.12e-02 & 92.59371 & 22.57190 & 966704178 & 202065204 & 0.54 & 3.83e-05 & 5.43e-03\\ 
65.55551 & 19.57757 & 780012100 & 210780789 & 1.33 & 1.13e-04 & 8.07e-03 & 93.65056 & 26.48868 & 105805190 & 202081077 & 0.45 & 1.34e-04 & 4.87e-03\\ 
66.12750 & 18.98697 & 966003188 & 210742017 & 2.88 & 6.22e-05 & 2.94e-02 & 97.33402 & 17.42132 & 960208224 & 202136166 & 1.14 & 1.52e-04 & 1.34e-02\\ 
66.45394 & 18.87997 & 420111044 & 210734946 & 0.42 & 1.04e-04 & 4.51e-03 & 134.08144 & 12.66383 & 796313013 & 211467243 & 0.26 & 5.70e-05 & 3.04e-03\\ 
66.54349 & 21.82035 & 680803069 & 210927181 & 0.25 & 1.24e-04 & 2.58e-03 & 138.67201 & 18.16239 & 200909023 & 211860434 & 0.71 & 1.21e-04 & 7.23e-03\\ 
67.17497 & 15.56479 & 369510103 & 210509151 & 1.10 & 2.16e-04 & 1.28e-02 & 138.73979 & 21.23177 & 1000410051 & 251351839 & 1.92 & 6.64e-05 & 2.26e-02\\ 
67.21062 & 18.74331 & 371413219 & 210725857 & 2.06 & 1.67e-04 & 1.86e-02 & 158.13676 & 6.50161 & 338114159 & 248624968 & 16.54 & 2.51e-04 & 1.63e-01\\ 
67.49813 & 24.55194 & 628810080 & 247820821 & 2.38 & 3.16e-04 & 1.90e-02 & 170.36677 & 2.08099 & 1085709096 & 201602470 & 0.94 & 1.03e-04 & 8.64e-03\\ 
67.50203 & 22.37514 & 680815230 & 247523445 & 0.90 & 1.92e-04 & 7.08e-03 & 172.02927 & 1.69179 & 237207028 & 201577109 & 0.17 & 7.53e-05 & 1.15e-03\\ 
67.51739 & 16.06889 & 1048610235 & 210540496 & 0.34 & 9.34e-05 & 3.66e-03 & 174.81451 & 7.16622 & 963605130 & 201892534 & 24.95 & 7.48e-06 & 2.95e-01\\ 
67.67754 & 14.66154 & 1048602207 & 246707008 & 1.20 & 4.47e-05 & 1.01e-02 & 175.39085 & -4.29261 & 499805062 & 201191890 & 0.39 & 8.32e-05 & 3.33e-03\\ 
67.81848 & 15.00324 & 282910030 & 246735080 & 1.07 & 3.77e-05 & 9.15e-03 & 177.02408 & 3.43871 & 794012161 & 201688096 & 1.16 & 7.75e-05 & 1.03e-02\\ 
67.85027 & 18.00590 & 371406142 & 210674635 & 2.21 & 1.84e-04 & 1.78e-02 & 177.72235 & 4.69415 & 457714004 & 201762168 & 0.26 & 1.05e-04 & 1.82e-03\\ 
68.07450 & 24.37072 & 759514229 & 247794636 & 1.73 & 7.91e-05 & 1.42e-02 & 182.79471 & 0.20064 & 200608235 & 201478655 & 26.14 & 8.29e-06 & 2.44e-01\\ 
68.09708 & 24.05038 & 680811209 & 247748412 & 3.37 & 4.40e-04 & 2.61e-02 & 344.68469 & -0.42519 & 367602089 & 246418383 & 0.37 & 1.75e-04 & 4.02e-03\\ 
68.10949 & 18.46449 & 976308147 & 247078637 & 2.57 & 4.78e-05 & 2.74e-02 & 286.51012 & 46.72143 & 908208033 & 9880223 & 3.05 & 9.17e-05 & 3.44e-02\\     
\enddata
\end{deluxetable*}

\section{Discussion}

From the LAMOST DR11 spectra of M stars, we have newly identified a total of 962 UCDs with spectral types later than M6 by cross-matching to the existed catalogs and a semi-supervised machine learning approach. The spectral types of our sample are visually checked. We have found that the spectra with a low resolution of 250 is sufficient for the identification of cool dwarfs and thus demonstrated that the CSST slitless survey has a potential to discover and characterize much more UCDs, at least for late-M to early-L dwarfs.

The generation of magnetic field in low-mass, fully convective dwarfs remains poorly understood. Combining with 734 UCDs previously identified by \citet{wang2022}, we have characterized the magnetic activity of the fully convective UCDs in the LAMOST survey through analyzing the H$\alpha$ emission. We have used the EWs of the H$\alpha$ line as the indicators of the chromospheric activity on UCDs. From the Kepler/K2 light curves, we have obtained the rotation periods for 82 targets with the Lomb-Scargle method. The activity-rotation relation seems to be saturated around Ro$_{\rm sat}$ of 0.1, more close to the value derived by \citet{douglas2014} and \citet{zhang2023} but smaller than those from \citet{newton2017} and \citet{li2024}. However, we only have a few of the sample with the Ro larger than 0.1. The lack of unsaturated sample limits our ability to constrain the activity-rotation relation. The convective turnover time of a ultracool dwarf with a stellar mass of 0.1$M_{\odot}$ is longer than 150 days and it increases with the decreased effective temperature. It is more challenging to derive long rotation periods from photometric surveys. \citet{west2015} found that all late-M dwarfs with the rotation periods shorter than 86 days are active.

\begin{figure}
\centering
\includegraphics[width=0.45\textwidth]{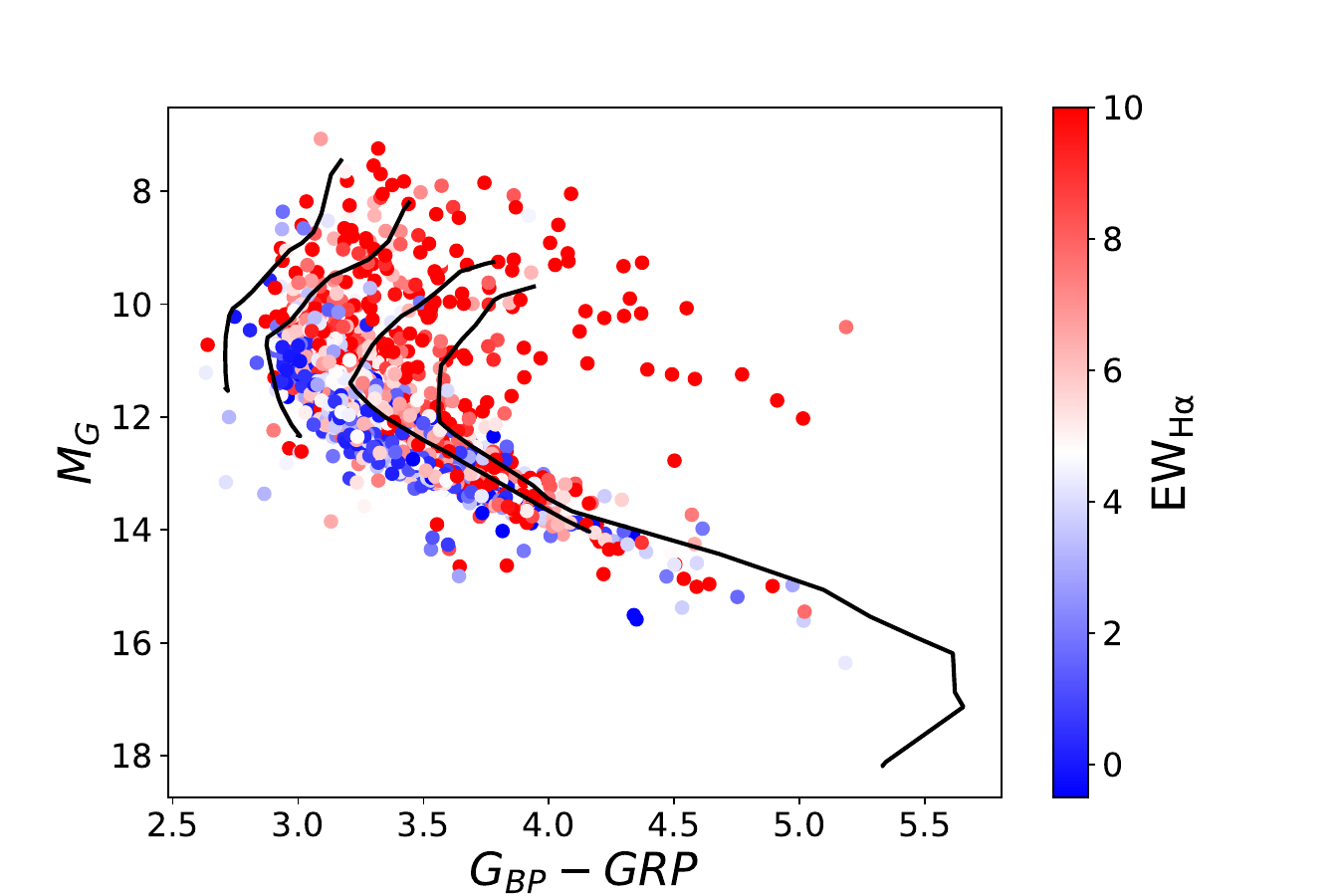}
\includegraphics[width=0.41\textwidth]{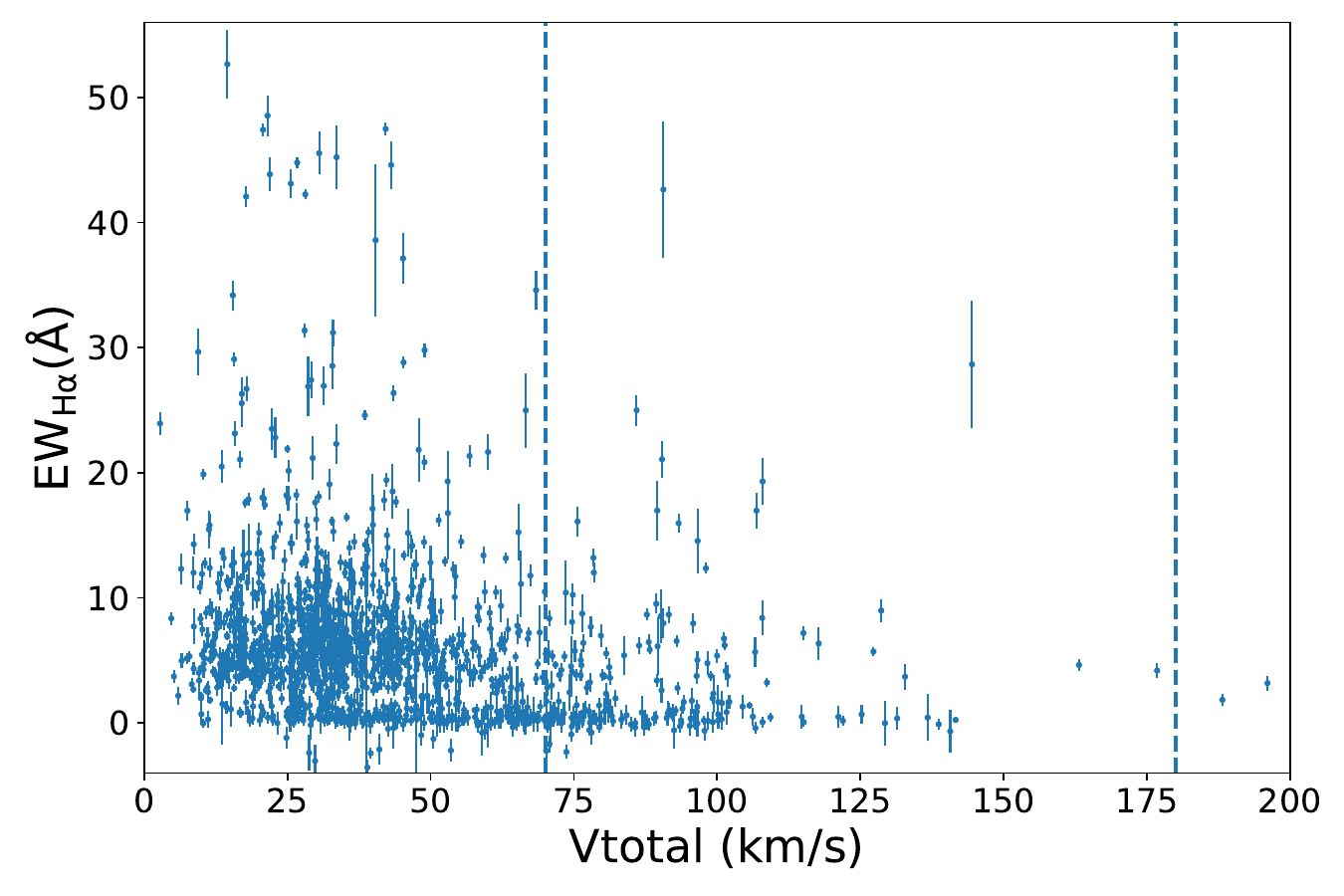}
\caption{The left panel shows the EWs of the H$\alpha$ line along the evolutionary track of \citet{baraffe2015}. The right panel shows the relation between the EWs of the H$\alpha$ line and the stellar kinematics.}
\label{fig:age}
\end{figure}

Precise determination of ages of low-mass UCDs is very difficult \citep{soderblom2010}, but we can get a crude estimate from the color-magnitude diagram with the model evolutionary track or the kinematic evolution. We plot the positions of our sample on color-magnitude diagram, along with the calculated EW$_{\rm H\alpha}$, see Figure \ref{fig:age}. The magnetic activity level of our sample decreases with age, probably due to the loss of rotational momentum. The young stars show higher levels of the H$\alpha$ emission as expected, which can be attributed to both of the strong magnetic activity and accretion process. On the other hand, the aging stars have larger velocities due to the dynamical heating. The right panel of Figure \ref{fig:age} shows how the EW$_{\rm H\alpha}$ relates to the total velocity, defined as $\sqrt{U^2+V^2+W^2}$. Young stars show stronger H$\alpha$ emission and lower stellar motions, located in the thin disk, while aged stars have much weaker H$\alpha$ emission and belong to the thick disk and thick disk-halo transition, consistent to results of \citet{west2006}.

\begin{figure}
\centering
\includegraphics[width=0.45\textwidth]{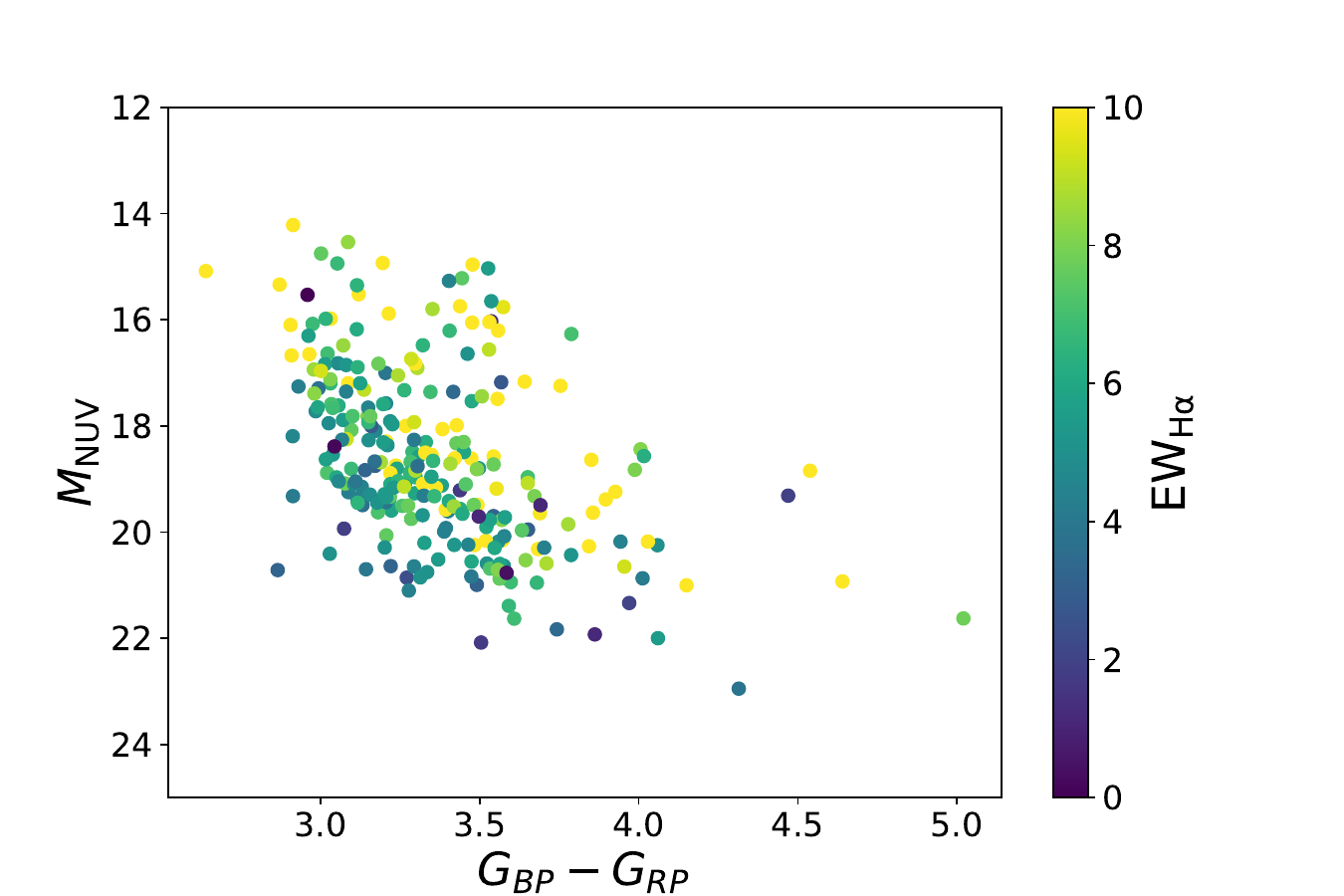}
\caption{The absolute NUV magnitude versus \br\ color diagram, as well as the EW$_{\rm H\alpha}$ shown in colors.}
\label{fig:nuv}
\end{figure}

Ultraviolet radiation is dominated by the chromosphere rather than the photosphere of an ultracool dwarf \citep{stelzer2013}. We have cross-matched the identified sample of UCDs with the GALEX observations to obtain the near-ultraviolet (NUV) magnitudes and then derive the absolute NUV magnitudes ($M_{\rm NUV}$) as the same manner for the $M_{G}$ using the Gaia parallaxes. We plot the relation of the NUV magnitude with $G_{BP}-G_{RP}$ color in Figure \ref{fig:nuv}, as well as the EW$_{\rm H\alpha}$ indicated by colors. In order to show more clearly, we restrict the upper limit of the colors to be the EW$_{\rm H\alpha}$ of 10\AA. The brightness of UCDs in the NUV band shows a large dispersion, which can be attributed to different level of chromospheric activity, and positively correlates to the H$\alpha$ line emission. The CSST's NUV bands of the photometry and spectroscopy are redder than that of the GALEX, but cover another important activity indicator, Mg II doublet around 2800\AA. The NUV observation ability of the CSST survey will be important for characterizing activity levels and (super)flares of nearby UCDs.

The magnitude of the slitless spectroscopic survey of the CSST will be in a range of 18-24, and is deeper by 1-2 magnitudes in the deep fields which have more observations. Due to the faintness of UCDs, the proportion of data that can be confirmed by the ground-based follow-up observations will be very small compared to the survey data. Hence, we think a semi-supervised approach will be useful in the CSST era. On the other hand, the CSST lacks the capability for infrared observations, where the peak energy of cool dwarfs and brown dwarfs is located. The Euclid satellite, which also performs a space-boned all-sky survey and highly overlaps the CSST observing areas, works in the near infrared (NIR) wavelength. The Euclid and CSST will be good complementary to each other in the future, making the observations to cover from the NUV to NIR wavelength.

In the near future, we will improve the performance and robust of the model to the real data with lower SNRs and errors in flux calibrations. Additionally, we will also incorporate the multi-band photometry, which has a deeper limiting magnitude, to not only detect new samples but also estimate their parameters.

\section{Conclusions}

In this work, we have identified UCDs from the LAMOST DR11 V1.0 and analyzed the magnetic activity-rotation relation based on both the LAMOST and Kepler/K2 data. We summarize this work as follows.

1. A total of 962 UCDs, including 774 M6, 137 M7, 41 M8, 10 M9 dwarfs, are newly identified from the LAMOST data through cross-matching with existing catalogs and a semi-supervised machine learning model. We demonstrate that the low-resolution CSST slitless spectra are capable of identifying UCDs. 

2. The H$\alpha$ line emissions of the UCDs are measured from the LAMOST spectra. More than 80 percent of our sample are active and the active fraction increases with spectral type.

3. We have found 82 UCDs that have both the LAMOST spectra and Kepler/K2 light curves and determined their rotational periods with the generalized Lomb-Scargle method. We derive the activity-rotation relation for these UCDs, which shows a saturation around the Rossby number of 0.12.\\

We are grateful to the anonymous referee for the valuable comments and suggestions. This work is supported by the National Natural Science Foundation of China under grant No. 12288102. The present study is also financially supported by the National Natural Science Foundation of China under grants Nos. 10373023, 10773027, U1531121, 11603068, 11903074 and 12373039, the Yunnan Fundamental Research Projects (grant Nos. 202201AT070186 and 202305AS350009), the Yunnan Revitalization Talent Support Program (Young Talent Project), International Centre of Supernovae, Yunnan Key Laboratory (No. 202302AN360001), and the China Manned Space Program with grant No. CMS-CSST-2025-A15. Guoshoujing Telescope (the Large Sky Area Multi-Object Fiber Spectroscopic Telescope LAMOST) is a National Major Scientific Project built by the Chinese Academy of Sciences. Funding for the project has been provided by the National Development and Reform Commission. LAMOST is operated and managed by the National Astronomical Observatories, Chinese Academy of Sciences. This paper includes data collected by the Kepler mission and obtained from the MAST data archive at the Space Telescope Science Institute (STScI). Funding for the Kepler mission is provided by the NASA Science Mission Directorate. STScI is operated by the Association of Universities for Research in Astronomy, Inc., under NASA contract NAS 5–26555. All the Kepler/K2 data used in this paper can be found in MAST: \dataset[10.17909/t98304]{https://doi.org/10.17909/t98304} and \dataset[10.17909/T9WS3R]{https://doi.org/10.17909/t9ws3r}.

\bibliographystyle{aasjournal}
\bibliography{ucdsdr11}

\end{CJK*}
\end{document}